\documentclass[12pt]{article}
\usepackage{graphicx}
\usepackage[utf8]{inputenc}
\usepackage[english
]{babel}
\def\baselinestretch{1.5}
\begin{document}
\begin{center}
\bf{Probability representation of quantum states as a renaissance of hidden variables -- God plays coins }\\
\end{center}
\smallskip

\begin{center} {\bf Vladimir N. Chernega$^1$, Olga V. Man'ko$^{1,2}$, Vladimir I. Man'ko$^{1,3}$}
\end{center}

\smallskip

\begin{center}
$^1$ - {\it Lebedev Physical Institute, Russian Academy of Sciences\\
Leninskii Prospect 53, Moscow 119991, Russia}\\
$^2$ - {\it Bauman Moscow State Technical University\\
The 2nd Baumanskaya Str. 5, Moscow 105005, Russia}\\
$^3$ - {\it Moscow Institute of Physics and Technology (State University)\\
Institutskii per. 9, Dolgoprudnyi, Moscow Region 141700, Russia}\\
Corresponding author e-mail: manko@sci.lebedev.ru
\end{center}

\smallskip

$^*$Corresponding author e-mail:~~~mankoov\,@\,lebedev.ru
\footnote[1]{Based on the invited talk presented at
a special session dedicated to Alexander S. Holevo's 75th birthday
of The International Conference ``Quantum Information, Statistics,
Probability'' (Steklov Mathematical Institute, Moscow,
September~12--15, 2018); [www.mathnet.ru/eng/conf1284].}

\begin{abstract}\noindent
We develop an approach where the quantum system states and quantum
observables are described as in classical statistical mechanics --
the states are identified with probability distributions and
observables, with random variables. An example of the spin-1/2 state
is considered. We show that the triada of Malevich's squares can be
used to illustrate the qubit state. We formulate the superposition
principle of quantum states in terms of probabilities determining
the quantum states. New formulas for nonlinear addition rules of
probabilities providing the probabilities associated with the
interference of quantum states are obtained. The evolution equation
for quantum states is given in the form of a kinetic equation for
the probability distribution identified with the state.
\end{abstract}

\medskip


\section{Introduction}  
The aim of our work is to demonstrate that in usual quantum
mechanics the notion of quantum system state can be formulated
analogously to the notion of classical system state in classical
statistical mechanics, namely, the quantum state can be identified
with the probability distribution. The classical system state, e.g.,
a state of the classical particle, is described by the particle
position $q$ and the particle velocity $\dot q$~\cite{Landau1} or
the particle momentum $p=m\,\dot q$, where $m$ is the particle mass.

If there are thermal fluctuations of the particle position and
momentum, the classical particle state is identified with the
probability density $f(q,p)$, which is a nonnegative function on the
phase-space plane with the normalization condition
$\displaystyle{\int} f(q,p)\,d q\,d p=1$. The other example of
classical system is a classical coin employed in the coin tossing
game.

If the coin is a nonideal one, its states corresponding to the
position ``up'' or ``down'' are described by the probability
distributions associated with the probability vectors $\vec
p=(p_1,p_2)$ with the nonnegative components $p_1$ and $p_2$
satisfying the normalization condition $p_1+p_2=1$. Here, $p_1$ is
the probability to have the coin position ``up'' and $p_2$ is the
probability to have the coin position ``down'' in the coin tossing
game.

For an ideal coin, the probabilities of the coin position ``up'' and
``down'' are equal, i.e., $p_1=p_2=1/2$. The positions ``up'' and
``down'' are random positions. We can identify the label ``up'' with
the number~1 and the position label ``down'' with the number~2. This
means that we can consider the coin probability distribution $p(j)$
as a function of random positions labeled by the integers (random
positions) $j=1,2$.

In the coin tossing game, we can consider the other random variables
given by functions $f(j)$, where $f(1)$ is identified with the gain
in the game, and $f(2)$ is identified with the loss in the game. The
classical statistics in the coin tossing game is associated with
random variable means $\langle f\rangle=p_1f(1)+p_2f(2)$ and other
highest moments $\langle
f^n\rangle=p_1f^n(1)+p_2f^n(2);\,n=2,3\ldots$

The coin state characteristics are given, e.g., by the Shannon
entropy $H=-p_1\ln p_1-p_2\ln
p_2$~\cite{Shannon,Holevobook,Holevo1,Khrennikov,Khrennikov1,Khrennikov4}.
There are other kinds of entropies, e.g., Tsallis
entropy~\cite{Tsallis} and R\'enyi entropy~\cite{Renyi}. In this
paper, we show that the same probability distributions of classical
coin positions determine these entropies for quantum states of
qubits.

If there are several coins, it is obvious that the states of these
coins are described by a set of the probability distributions
$p_j^{(n)}$, where $j=1,2$ and $n=1,2,\ldots N$. The label $n$
corresponds to the $n$th coin probability distribution. If there is
no correlations among the coins, the probabilities $p_1^{(n)}$ and
$p_2^{(n)}$ satisfy the only constraints $0\leq
p_1^{(n)},\,p_2^{(n)}\leq 1$ and $p_1^{(n)}+p_2^{(n)}=1$.

In the case of a dependence of coin positions, extra relations can
exist for numbers $p_1^{(n)}$ and $p_2^{(n)}$. For example, if two
coins are completely identical, we have $p_j^{(1)}=p_j^{(2)}$; this
means that one has the possibility to consider the probability
distributions for two coins, where the distribution for second coin
is completely determined by the probability distribution for the
first coin, being simply the same distribution.

In quantum mechanics, Schr\"odinger~\cite{Schrodinger} introduced
the notion of the particle state (pure state), identifying it with
the complex wave function $\psi(q)=|\psi(q)| \exp[i\phi(q)]$. The
modulus of the wave function has the intuitively clear physical
meaning of the probability density $P_\psi(q)=|\psi(q)|^2$, but the
phase of the complex wave function $\phi(q)$ does not have an
analogous probabilistic interpretation.

The influence of thermal fluctuations was taken into account by
considering mixed states of the particle. Landau~\cite{Landau} and
von Neumann~\cite{vonNeumann} introduced the notion of mixed states
identifying the particle state with complex Hermitian density matrix
$\rho(q,q')$, which is a function of two variables with the property
$\rho(q,q')=\rho^\ast(q',q).$ For normalized states
$\displaystyle{\int}\rho(q,q)\,d q=1$ and for pure states with wave
function $\psi(q)$, the density matrix is expressed in terms of the
wave function $\rho_\psi(q,q')=\psi(q)\psi^\ast(q')$. This picture
was generalized by Dirac, and the pure states were identified with
the state vectors $|\psi\rangle$, and the mixed states were
identified with the density operators $\hat\rho$, where the vectors
live in the Hilbert space, and operators $\hat\rho$ act in the
Hilbert space~\cite{Diracbook}.

One can see that the described notion of quantum system states is
drastically different from the considered notion of states in
classical statistics. In view of this, from the early days of
quantum mechanics there were attempts to find the formulation of
quantum mechanics with the possibility to associate the states with
other functions, which are similar to functions used in classical
statistics. The first attempt was made by Wigner~\cite{Wig32}, who
introduced the Wigner function $W(p,q)$, which is similar to the
classical probability distribution $f(q,p)$ but this function is not
the probability density since it can take negative values.

Later Husimi~\cite{Husimi40} and Kano~\cite{Kano56} introduced the
function $Q(q,p)$, which has only nonnegative values, but its
continuous variables $q$ and $p$ are not physical observables
(position and momentum), since the uncertainty
relations~\cite{Heisenberg1927,Schrodinger1930,Robertson1930}
prohibit one from measuring simultaneously the position and
momentum. This means that a joint probability distribution of two
random continuous position and momentum does not exist. In view of
this fact, the Wigner and Husimi--Kano functions are called
quasidistributions.

Other kinds of quasidistributions $P(q,p)$ were introduced by
Glauber~\cite{Glauber63} and Sudarshan~\cite{Sudarshan63} and by
Blokhintsev~\cite{Blohintsev,Blohintsev1}. All these
quasidistributions are related by different integral transforms with
the density matrix $\rho(q,q')$. They are different representations
of the density operator $\hat\rho$ determining the state of the
particle. In~\cite{Mancini96}, the notion of the quantum particle
state was introduced, which identifies it with the fair probability
density $w(X,\mu,\nu)$ of a random variable $X$ and parameters $\mu$
and $\nu$, called a symplectic tomogram, which is a nonnegative
function satisfying the normalization condition $\displaystyle{\int
}w(X,\mu,\nu)\,dX=1$.

A symplectic tomogram is a generalization of the optical tomogram
$w(X,\theta)$~\cite{BerBer,VogRis} used in
experiments~\cite{RaymerPRL93} as a technical instrument to measure
the photon state identified with the Wigner function. The idea
of~\cite{Mancini96,ManciniFoundPhys} was to suggest the
identification of the quantum-particle-state notion with standard
probability density. It was shown in~\cite{my1} that the notion of
symplectic tomogram $w(X,\mu,\nu)$ and optical tomogram
$w(X,\theta)$ can be also introduced in classical statistics, where
these probability densities are expressed in terms of the
probability density $f(q,p)$ by means of the invertible integral
Radon transform~\cite{Radon1917}.

In classical statistics, this transform maps the probability density
$f(q,p)$ onto the tomographic pro\-bability
density~\cite{Pilyavets}. Interestingly, in quantum mechanics
exactly the same invertible transform maps the Wigner
quasidistribution $W(q,p)$, which can take negative values, onto a
fair nonnegative tomographic probability density. Reviews of this
approach are given
in~\cite{Asorey,AmosovRorennoyPRA,KorennoyMankoJRLR2011,ChernegaJETPLett}.
Examples of using the tomographic probability distributions for
oscillator systems were presented
in~\cite{OVMDampedscilKRPen,OVNVJRLR,VRMB}.

It turned out that the probability distribution determining the spin
states (qudit states, $N$-level atom states) can be also
introduced~\cite{DodPLA,OlgaJetp,Bregence,PainiDariano,Weigert,Wiegert2}.
There exist the bijective maps of density matrices of spin states
onto standard probability distributions both of tomographic
probability
type~\cite{TMFSafonov,AndreevJRLR1998,ActaPhysHun,FillJRLR} and the
coin tossing game
type~\cite{Chernega1,Chernega2,Chernega3,Chernega4,Chernega5,ChernegaEJP,Entropy1,Entropy2,Entropy3,MarmoVitaleJPA}.
In the case of qudit states, the quantum observables (Hermitian
matrices) can be mapped bijectively onto sets of classical-like
dichotomic
variables~\cite{Chernega1,Chernega2,Chernega3,Chernega4,Chernega5,ChernegaEJP}.

The aim of the paper is to present a review of the suggested
probability representation of quantum system states, including the
superposition principle of the states expressed in terms of the new
nonlinear addition rule of the probabilities~\cite{Chernega3} and
illustration of the qubit states in quantum suprematism
picture~\cite{Chernega5,ChernegaEJP}, where the Bloch sphere
parameters of the states are bijectively mapped onto the triada of
Malevich's squares (red, black, and white) associated with triangle
geometry of these
states~\cite{Chernega1,Chernega2,Chernega3,Chernega4,Chernega5,ChernegaEJP,Entropy2,Entropy3}.

This paper is organized as follows.

In Sec.~2, the elements of complex matrix are expressed in terms of
the probabilities. In Sec.~3, the bijective map for classical random
variables used in the coin tossing game and Hermitian operators
acting in the Hilbert space are constructed. In Sec.~4, the
superposition principle in the probability representation is
reviewed. In Sec.~5, the unitary evolution of the density matrix is
expressed in terms of probability distribution transforms. In
Sec.~6, the notion of quantum suprematism is discussed, and the main
results of this work are pointed out in Sec.~7.

\section{Matrices as Probability Distribution}
We discuss the map of the generic matrices onto the probability
distributions. Such map seems to have not been considered in the
literature. In~\cite{Sud}, the matrices $A$ were bijectively mapped
onto vectors $|A\rangle$, using the rule illustrated by the example
of 2$\times$2 matrices
\begin{equation} \label{eq.1}
A=\left(\begin{array}{cc}
  A_{11}&A_{22}\\
  A_{21}&A_{22}
\end{array}\right)~\leftrightarrow~
|A\rangle=\left(\begin{array}{c}
  A_{11}\\
  A_{12}\\
  A_{21}\\
  A_{22}
\end{array}\right).
\end{equation}
Thus, the matrix elements in the rows of the matrix $A$ construct
the column of the vector $|A\rangle$. The 4$\times$4~matrix
$A$$\times$$ A^\ast$ is bijectively mapped onto the 16-dimensional
vector $|A'\rangle$ with components $A_1=|A_{11}|^2,$
$A_{2}=A_{11}A^\ast_{12}$, $A_3=A_{12}A^\ast_{11}$,
$A_4=|A_{12}|^2,$ $A_5=A_{11}A_{21}^\ast,$ $A_6=A_{11}A_{22}^\ast$,
$A_7=A_{12}A_{21}^\ast$, $A_8=A_{12}A_{22}^\ast$, $
A_9=A_{21}A_{11}^\ast,$ $A_{10}=A_{21}A_{12}^\ast$,
$A_{11}=A_{22}A_{11}^\ast$, $A_{12}=A_{22}A_{12}^\ast$,
$A_{13}=|A_{21}|^2$, $A_{14}=A_{21}A_{22}^\ast$,
$A_{15}=A_{22}A_{21}^\ast$, and $A_{16}=|A_{22}|^2.$

The matrix $|A\rangle\langle A|$ is mapped onto the other vector
with 16~components, namely, the vector $|\tilde A\rangle$. Its
components are $\tilde A_1=|A_{11}|^2,$ $\tilde
A_{2}=A_{11}A^\ast_{12}$, $\tilde A_3=A_{11}A^\ast_{21}$, $\tilde
A_4=A_{11}A_{22}^\ast,$ $\tilde A_5=A_{12}A_{11}^\ast,$  $\tilde
A_6=|A_{12}|^2t$, $\tilde A_7=A_{12}A_{21}^\ast$, $\tilde
A_8=A_{12}A_{22}^\ast$, $\tilde A_9=A_{21}A_{11}^\ast,$ $\tilde
A_{10}=A_{21}A_{12}^\ast$, $\tilde A_{11}=|A_{21}|^2$, $\tilde
A_{12}=A_{21}A_{22}^\ast$, $\tilde A_{13}=A_{22}A_{11}^\ast $,
$\tilde A_{14}=A_{22}A_{12}^\ast$, $\tilde
A_{15}=A_{22}A_{21}^\ast$, and $\tilde A_{16}=|A_{22}|^2.$

The vectors $|A'\rangle$ and $|\tilde A\rangle$ are connected by the
unitary transform
\begin{equation}
|A'\rangle = T|\tilde A\rangle ,
\end{equation}
where the 16$\times$16~matrix $T$ has the block form $T_{ik}$ with
unity and zero 2$\times$2 blocks $(i, k=1,2,\ldots,8)$. The nonzero
blocks are
$T_{11}=T_{13}=T_{31}=T_{44}=T_{55}=T_{67}=T_{76}=T_{88}=1_2$. The
matrix $$
\rho = \frac{|\tilde A\rangle\langle\tilde A|}{\langle\tilde A|\tilde A\rangle}
$$ is the Hermitian matrix, and its trace is equal to unity. The
eigenvalues of this matrix are equal either to zero or unity, i.e.,
$\rho^2=\rho$ and $\mbox{Tr}\rho^2=\mbox{Tr}\rho=1$.

As it was conjectured
in~\cite{Chernega5,ChernegaEJP,Entropy1,Entropy2,Entropy3}, an
arbitrary 4$\times$4~matrix $\rho$ with such properties (it can be
interpreted as the density matrix of the pure state of spin-3/2
system or of two-qubit systems) has the probability parametrization,
i.e., the matrix is
\begin{equation} \label{eq.3}
\rho = \left(\begin{array}{cccc}
  \rho_{11}&\rho_{12}&\rho_{13}&\rho_{14}\\
  \rho_{21}&\rho_{22}&\rho_{23}&\rho_{24}\\
  \rho_{31}&\rho_{32}&\rho_{33}&\rho_{34}\\
  \rho_{41}&\rho_{42}&\rho_{43}&\rho_{44}
\end{array}\right),
\end{equation}
where nonnegative numbers
\begin{equation} \label{eq.4}
\rho_{22} = 1-p_{3}^{(22)},\quad\rho_{33}=1-p_{3}^{(33)},\quad
\rho_{44} = 1-p_{3}^{(44)},\quad\rho_{11}=\rho_3^{(22)}+\rho_3^{(33)}+p_3^{(44)}-2
\end{equation}
provide the diagonal matrix elements of the matrix $\rho$ in terms
of probabilities $0\geq p_3^{(22)}\geq1$, $0\geq p_3^{(33)}\geq1$,
and $0\geq p_3^{(44)}\geq 1$ associated with three probability
distributions $(p_3^{(22)},1-p_3^{(22)})$,
$(p_3^{(33)},1-p_3^{(33)})$, and $(p_3^{(44)},1-p_3^{(44)})$
describing the tossing coin game for three coins labeled by indices
$(22)$, $(33)$, and $(44)$. The nondiagonal elements $\rho_{j
k},\,j>k$ of the matrix $\rho$ read
\begin{equation} \label{eq.5}
\rho_{j k} = p_{1}^{(j k)}-(1/2)+i(p_2^{(j k)}-1/2),\qquad
j,k=2,3,4,\quad j>k,
\end{equation}
where the nonnegative numbers $0\leq p_1^{(j k)}\leq 1$ and $0\leq
p_2^{(jk)}\leq 1$ are associated with the probability distributions
$(p_1^{(j k)},1-p_1^{(jk)})$ and $(p_2^{(jk)},1-p_2^{(jk)})$
describing the tossing coin gains for 12~coins labeled by the
indices $(jk)$.

The probabilities satisfy the inequalities of nonnegativity of the
matrix $\rho\geq 0$. If the matrix $A$ satisfies the condition
$\mbox{Tr}(A^\dagger A)=1$, its matrix elements can be connected with the
probabilities $p_{1,2,3}^{(jk)}$; $j,k=1,2,3,4$ as follows:
\begin{eqnarray}
  &&|A_{11}|^2 = p_3^{(22)}+p_3^{(33)}+p_3^{(44)}-2,\nonumber\\
&& A_{11}A_{12}^\ast=p_1^{(12)}
  -(1/2)-i(p_2^{(12)}-1/2)=x,\nonumber\\
  && A_{11}A_{21}^\ast = p_1^{(13)}-(1/2)-i(p_2^{(13)}-1/2)=y,\nonumber\\
&&A_{11}A_{22}^\ast=p_1^{(14)}-(1/2)-i(p_2^{(14)}-1/2)=z,\nonumber\\
  && |A_{12}|^2 = 1-p_3^{(22)},\quad
  A_{12}A_{21}^\ast=p_1^{(23)}-(1/2)-i(p_2^{(23)}-1/2)=m,\nonumber\\[-2mm]
  \label{eq.6}\\[-2mm]
  && A_{12}A_{22}^\ast = p_1^{(24)}-(1/2)-i(p_2^{(24)}-1/2)=n,\quad
|A_{21}|^2 = 1-p_3^{(33)},\nonumber\\
  && A_{21}A_{22}^\ast=p_1^{(3 4)}-(1/2)-i(p_2^{(3 4)}-1/2)=t,\quad
|A_{22}|^2 = 1-p_3^{(44)},\nonumber\\
  && |A_{11}|^2+|A_{21}|^2+|A_{12}|^2+|A_{22}|^2=1.\nonumber
\end{eqnarray}
Relations (\ref{eq.6}) provide the possibility to express the
complex matrix elements $A_{11},\,A_{12},\,A_{21},\,A_{22}$ given by
seven parameters of the matrix $A_{j k}$; $j,k=1,2$ in terms of the
above probabilities $p_{1,2}^{(jk)}$, where $j,k=1,2,3,4$, and
$p_3^{(jj)}$, where $j=2,3,4.$

We demonstrated that for the matrix $A$ one can introduce the
probability representation. An analogous construction can be
introduced for $N$$\times $$N$~matrix $A$, where $N=3,4,\ldots$

\section{Qubit States as the Quantization of Classical Coin
Probability Distributions}  
As we demonstrated in the previous section, the
matrix elements of an arbitrary matrix $A$ can be related to some
probability distributions. We use this observation to suggest the
following quantization procedure of classical statistics. We
construct from classical probability distributions, describing the
coin states, the density matrices and state vectors in a Hilbert
space. This means that we map the probability
distributions~(simplexes) onto density operators acting in the
Hilbert space. Also we map the classical random variables, used in
the coin tossing game, onto Hermitian matrices, i.e., Hermitian
operators acting in the Hilbert space. The map constructed is a
bijective map.

After the map was constructed, we obtain all the relations known in
quantum mechanics rewritten as the relations for classical-like
random variables and standard probability distributions. We
demonstrate this procedure on the example of spin-1/2 (qubit,
two-level atom) state and spin observables.

Let us start with tossing three nonideal classical coins. The states
of these three coins are identified with three probability
distributions with probabilities $0\leq p_1,p_2,p_3\leq 1$, i.e.,
$(p_1,1-p_1)$, $(p_2,1-p_2)$, and $(p_3,1-p_3)$. The probabilities
$p_1$, $p_2$, and $p_3$ are the probabilities in the game to have
for each coin the position ``up.''

As we discussed in the introduction, the dichotomic random variables
$f^{(n)}(j)$; $j=1,2;\,n=1,2,3$ associated with these three coins
can be denoted as
$f^{(1)}(1)=x;\,f^{(1)}(2)=-x,\,f^{(2)}(1)=y;\,f^{(2)}(2)=-y$, and
$f^{(3)}(1)=z_1;\,f^{(3)}(2)=z_2$, where $x,y,z_1$, and $z_2$ are
real numbers. The statistics of the coin tossing game provides mean
values of random variables, such as
\begin{equation} \label{eq.1a}
\langle f^{(1)}\rangle=x p_1-x(1-p_1), \quad
  \langle f^{(2)}\rangle=y p_2-y(1-p_2),\quad
  \langle f^{(3)}\rangle=p_3z_1+(1-p_3)z_2.
\end{equation}
The highest moments of the classical random variables are given by
standard formulas of the probability theory ($k=2,3,\ldots$) as
follows:
\begin{equation} \label{eq.2a}
\langle f^{(1)k}\rangle = x^k p_1+(-x)^k(1-p_1), \quad
  \langle f^{(2)k}\rangle=y^k p_2+(-y)^k(1-p_2), \quad
  \langle f^{(3)k}\rangle=p_3z_1^k+(1-p_3)z_2^k.
\end{equation}
Let us organize the numbers $p_1$, $p_2$, and $p_3$ in table form
given as the 2$\times$2~matrix
\begin{equation} \label{eq.3a}
\rho = \left(\begin{array}{cc}
  p_3 & p_1-1/2-i(p_2-1/2)\\
  p_1-1/2+i(p_2-1/2) & 1-p_3\end{array}\right).
\end{equation}
The states of classical three coins can be associated with matrix
elements of the Hermitian trace-one matrix $\rho$.

Now we introduce the quantization procedure; namely, we impose the
condition for this Hermitian matrix $\rho$ to have only nonnegative
eigenvalues. This means that $\mbox{det}\rho\geq0$ or the
probabilities $p_1$, $p_2$, and $p_3$ satisfy the condition
\begin{equation} \label{eq.4a}
(p_1-1/2)^2+(p_2-1/2)^2+(p_3-1/2)^2 \leq 1/4.
\end{equation}
Comparing matrix~(\ref{eq.3a}) with density matrices of spin-1/2
states, we see that matrix~(\ref{eq.3a}) simulates all the possible
quantum states of this system. For the case $\rho^2=\rho$, we obtain
the condition for pure spin-1/2 states
\begin{equation} \label{eq.5a}
(p_1-1/2)^2+(p_2-1/2)^2+(p_3-1/2)^2=1/4.
\end{equation}
In this case, we can construct the vector $|\psi\rangle$ in the
Hilbert space, which provides the matrix elements of the matrix
$\rho$. One can check that the vector  $|\psi\rangle$ in Dirac
notation reads
\begin{equation} \label{eq.6a}
|\psi\rangle = \left(\begin{array}{c}
  \sqrt{p_3}\\[2mm]
   \frac{p_1-(1/2)+i(p_2-1/2)}{\sqrt{p_3}}\end{array}\right).
\end{equation}
In fact, $|\psi\rangle\langle\psi|=\rho_\psi$. We constructed this
vector by introducing a triple of normalized vectors of the basis in
the linear space ${\cal H}$ (the Hilbert space) of the form
\begin{eqnarray}
  &&|1\rangle_z=\left(\begin{array}{c}
  1\\
  0
  \end{array} \right),\quad
|2\rangle_z = \left(\begin{array}{c}
  0\\
  1
\end{array}\right),\quad
|1\rangle_x = \frac{1}{\sqrt2}\left(\begin{array}{c}
  1\\
  1
\end{array}\right),\nonumber\\[-2mm]
&&\label{eq.7a}\\[-2mm]
&&|2\rangle_x =
\frac{1}{\sqrt2}\left(\begin{array}{c}
  1\\
  -1
\end{array}\right),\quad
|1\rangle_y=\frac{1}{\sqrt2}\left(\begin{array}{c}
  1\\
  i
\end{array}\right),\quad
|2\rangle_y=\frac{1}{\sqrt2}\left(\begin{array}{c}
  1\\
  -i
\end{array}\right).\nonumber
\end{eqnarray}
We introduced these special basis vectors (not yet considered in the
literature of the usual probability theory), since they provide the
matrices
\begin{eqnarray}
&&\rho_x^{(1/2)} = (|1\rangle_x\,_x\langle1|)=\frac{1}{2}\left(
  \begin{array}{cc}
  1 & 1\\
  1 & 1\end{array}\right),\qquad
\rho_x^{(-1/2)} = (|2\rangle_x\,_x\langle2|)=\frac{1}{2}\left(
  \begin{array}{cc}
  1 & -1\\
  -1 & 1\end{array}\right),\nonumber\\
&&\rho_y^{(1/2)} = (|1\rangle_y\,_y\langle1|)=\frac{1}{2}\left(
  \begin{array}{cc}
  1 & -i\\
  i & 1\end{array}\right),\qquad
\rho_y^{(-1/2)} = (|2\rangle_y\,_y\langle2|)=\frac{1}{2}\left(
  \begin{array}{cc}
  1 & i\\
  -i & 1\end{array}\right),\label{eq.8a}\\
&&\rho_z^{(1/2)} =
\frac{1}{2}|1\rangle_z\,_z\langle1|=\frac{1}{2}\left(
  \begin{array}{cc}
  1 & 0\\
  0 & 0\end{array}\right),\qquad
\rho_z^{(-1/2)}=\frac{1}{2}|2\rangle_z\,_z\langle2|=\frac{1}{2}\left(
  \begin{array}{cc}
  0 & 0\\
  0 & 1\end{array}\right). \nonumber
\end{eqnarray}
The Hermitian matrices (\ref{eq.8a}) are related to the classical
random variables $x,\, y,\, z_1$, and $z_2$. In fact, one can
organized these random variables in the form of Hermitian matrix
\begin{equation} \label{eq.9a}
H = \left(\begin{array}{cc}
  z_1 & x-i y\\
  x+i y & z_2\end{array}\right).
\end{equation}
These matrices can be expressed in terms of the identity matrix {\bf
1} and Pauli matrices
\begin{equation} \label{eq.10a}
\sigma_x = \left(\begin{array}{cc}
  0 & 1\\ 1 & 0 \end{array}\right),\qquad
\sigma_y = \left(\begin{array}{cc}
  0 & -i\\ i & 0 \end{array}\right),\qquad
\sigma_z = \left(\begin{array}{cc}
  1 & 0\\ 0 & -1 \end{array}\right).
\end{equation}
One has
\begin{equation} \label{eq.11a}
H = \frac{1}{2}\left[(z_1+z_2){\bf 1}+(z_1-z_2)\sigma_z+x\sigma_x
  +y\sigma_y\right].
\end{equation}
Expressions (\ref{eq.8a})--(\ref{eq.11a}) show that an arbitrary
quantum spin-1/2 observable presented by the Hermitian matrix $H$
can be simulated by three classical dichotomic random variables
taking values $(x,-x)$, $(y,-y)$, and $(z_1,z_2)$. For example, the
spin-1/2 projection $s_x=\sigma_x/2,$ $s_y=\sigma_y/2$, and
$s_z=\sigma_z/2$ are simulated by classical random variables;
namely, for the $s_x$ observable, the random variables are
$z_1=z_2=0$, $x=1/2$, $-x=-1/2$, and $y=0$; for the $s_y$
observable, $z_1=z_2=0$, $x=0$, $y=1/2$, and $-y=-1/2$; for the
$s_z$ observable, $x=0$, $y=0$, $z_1=1/2$, and $z_2=-1/2$. Thus, for
the coin tossing game, the conditions associated with spin
projections onto three perpendicular directions give observables
corresponding to the three coins, and they are equivalent for the
directions of axes $x,$ $y$, and $z$.

If we organize the classical random variables in the form of
Hermitian matrices, the different observables do not commute. For
spin projections, the quantum observables $s_x$, $s_y$, and $s_z$
provide the commutation relation
\begin{equation} \label{eq.9a}
[s_x,s_y] = (s_x s_y-s_y s_x) = i s_z.
\end{equation}
As well as the condition (\ref{eq.4a}), this relation means that our
quantization procedure imposes correlations for classical coin
states and observables. The quantization condition means also that,
if we impose the formula for statistical properties of the quantum
observable $H$ of standard form
\begin{equation} \label{eq.10a}
\langle H^k\rangle = \mbox{Tr}(\rho H^k),\, k=1,2,\ldots,
\end{equation}
the quantum statistics can be related to classical probabilities
simulating the state $\rho$ and classical random dichotomic
variables simulating the quantum observable $H$. For example, the
mean value of $H$ is
\begin{equation} \label{eq/11a}
\langle H\rangle = x(2p_1-1)+y(2p_2-1)+p_3(z_1-z_2)+z_2.
\end{equation}
The quantization procedure of classical random variables by means of
writing the variables $x,$ $y$, $z_1$, and $z_2$ in the matrix
form~(15) means the imposing the condition that the eigenvalues of
the matrix $H$ are the values that provide the results of the
measurement of quantum observable~(17) in the state with density
matrix~(\ref{eq.3a}). Thus, the quantization procedure gives the
possible values of quantum observable eigenvalues $H_1$ and $H_2$
expressed in terms of three classical random variables of the form
\begin{equation} \label{eq.11aa}
H_1 = \frac{z_1+z_2}{2}+\sqrt{\frac{(z_1-z_2)^2+x^2+y^2}{4}},\qquad
H_2 = \frac{z_1+z_2}{2}-\sqrt{\frac{(z_1-z_2)^2+x^2+y^2}{4}}.
\end{equation}
We also point out that the eigenvalues $\lambda_1$ and $\lambda_2$
of the density matrix $\rho$~(\ref{eq.3a}) are the probabilities
expressed in terms of the probabilities $p_1$, $p_2$, and $p_3$ as
follows:
\begin{eqnarray}    
&&\lambda_1 = \frac{1}{2}+\left\{\frac{1}{4}-\left[p_3(1-p_3)-(p_1-1/2)^2
  -(p_2-1/2)^2\right]\right\}^{1/2},\nonumber\\[-2mm]
&&\label{eq.12a}\\[-2mm]
&&\lambda_2 =
\frac{1}{2}-\left\{\frac{1}{4}-\left[p_3(1-p_3)-(p_1-1/2)^2
  -(p_2-1/2)^2\right]\right\}^{1/2}.\nonumber
\end{eqnarray}
The eigenvectors of the matrix $\rho$ read
\begin{eqnarray}    
&&|\rho_1\rangle = \left[1+\frac{\lambda_1-p_3}{(p_1-1/2)^2+(p_2-1/2)^2}\right]^{-1/2}
  \left(\begin{array}{c}
  1  \\
  \frac{\lambda_1-p_3}{(p_1-1/2)-i(p_2-1/2)}\end{array}\right),\nonumber\\[-2mm]
\label{eq.12'a}\\[-2mm]
&&|\rho_2\rangle=\left[1+\frac{\lambda_2-p_3}{(p_1-1/2)^2+(p_2-1/2)^2}\right]^{-1/2}
  \left(\begin{array}{c}
  1 \\ \frac{\lambda_2-p_3}{(p_1-1/2)-i(p_2-1/2)}\end{array}\right).\nonumber
\end{eqnarray}
The eigenvectors of the matrix $H$ determine the mean values of
quantum observables $H$ in a given state with the density
matrix~(\ref{eq.3a}). In fact, the unitary matrix $U$, such that
\begin{eqnarray} \label{eq.13a}
H = \left(\begin{array}{cc}
  u_{11} & u_{12}\\
  u_{21} & u_{22}\end{array}\right)
\left(\begin{array}{cc}
  H_{1} & 0 \\
  0 & H_2 \end{array}\right)
\left(\begin{array}{cc}
  u^\ast_{11} & u^\ast_{12}\\
  u^\ast_{21} & u^\ast_{22}\end{array}\right),
  \end{eqnarray}
defines the bistochastic matrix
\begin{equation} \label{eq.14a}
M = \left(\begin{array}{cc}
  |u_{11}|^2 & |u_{12}|^2\\
  |u_{21}|^2 & |u_{22}|^2\end{array}\right),
\end{equation}
where the probability
\begin{equation} \label{eq.15a}
|u_{11}|^2 = \left[1+\frac{|H_1-z_1|^2}{x^2+y^2}\right]^{-1/2}
\end{equation}
relates the mean value $H_{11}$ of the observable $H$ with the
eigenvalues (21) as follows:
\begin{eqnarray}
H_{11} = |u_{11}|^2H_1+(1-|u_{11}|^2)H_2,\label{eq.16a}\\
H_{22} = |u_{11}|^2H_2+(1-|u_{11}|^2)H_1,\label{eq.17a}
\end{eqnarray}
Formulas (21), (22), and (\ref{eq.15a})--(\ref{eq.17a}) provide the
expressions of quantum eigenvalues and eigenvectors both for
observables $H$ and density matrix $\rho$ in terms of classical coin
probabilities $p_1,p_2$, and $p_3$ and classical observables
$x,y,z_1$, and $z_2$.

The von Neumann entropy  $S=-\mbox{Tr}\rho\ln\rho$   of the quantum state
with density matrix~(\ref{eq.3a}) is expressed in terms of classical
coin probabilities $p_1,p_2$, and $p_3$, in view of
Eq.~(\ref{eq.12a}),
\begin{equation} \label{eq.18a}
S = -\lambda_1\ln\lambda_1-\lambda_2\ln\lambda_2.
\end{equation}
Also the Tsallis entropy of quantum states can be expressed in terms
of classical probabilities as
\begin{equation} \label{eq.19a}
S_T = \frac{\lambda_1^q+\lambda_2^q-1}{1-q},
\end{equation}
where $\lambda_1$ and $\lambda_2$ depend on the coin position
probabilities $p_1,p_2$, and $p_3.$

In view of our approach, we can formulate a problem analogous to the
one studied by Koopman~\cite{Koopman} and von
Neumann~\cite{vonNeumann1}. The initial idea was to formulate
quantum mechanics using the formalism of classical statistics. The
inverse problem considered in~\cite{Koopman,vonNeumann1} was to
formulate classical statistical mechanics using the formalism of
quantum mechanics like a Hilbert space and operators acting in the
Hilbert space. Some examples of this approach are related to
constructing the wave function of classical
oscillator~\cite{my2wavefunc,Vovaoscilwavefun,AshotVI}.

The problem that we are now considering is an analogous one. Namely,
it is possible that the standard probability theory or the theory of
simplexes is mapped onto the formalism of Hilbert spaces and
operators acting in the Hilbert spaces. We see that in the usual
probability theory the constructions induced by the relations
available in quantum mechanics appear. These constructions seem not
to have been considered in the literature connected with the
probability theory like pseudostochastic matrices~\cite{GrushiMan},
transforming probability distributions with constraints related to
the density-matrix properties. Such kinds of probability-theory
problems can be solved using the solutions of quantum mechanics
problems in the probability representation of quantum states and
sequent employment of the map of discussed properties of states and
observables onto probability distributions and random variables.

\section{Superposition Principle in the Probability Representation} 
Since the pure qubit states are expressed in terms of probabilities,
one can obtain the formulas for the superposition of states
$|\psi_1\rangle$ and $|\psi_2\rangle$ in the form of addition of the
probabilities. To obtain these formulas, we use the
results~\cite{Sud} where the superposition of two pure states with
density operators $\hat\rho_1=|\psi_1\rangle\langle\psi_1|$ and
$\hat\rho_2=|\psi_2\rangle\langle\psi_2|$ was expressed as a
nonlinear addition of the density operators giving the projector
$\hat\rho_\psi$, which is
\begin{equation} \label{eq.1b}
\hat\rho_\psi = \lambda_1\hat\rho_1+\lambda_2\hat\rho_2
  +\sqrt{\lambda_1\lambda_2}\,\frac{\hat\rho_1\hat\rho_0\hat\rho_2
  +\hat\rho_2\hat\rho_0\hat\rho_1}{\sqrt{\mbox{Tr}(\hat\rho_1\hat\rho_0\hat\rho_2\hat\rho_0)}}.
\end{equation}
Here, $0\leq\lambda_1$, $\lambda_2\leq1$, and
$\lambda_1+\lambda_2=1$ are the probabilities, and $\hat\rho_1$,
$\hat\rho_0$, and $\hat\rho_2$ are given pure-state density
operators written in the probability representation. Expressing the
pure state density matrices $\rho_\psi$, $\rho_1,$ ,$\rho_2$, and
$\rho_0$ in terms of the probabilities, we derive the addition rule
of probabilities~\cite{Chernega3,MAVI2019}  corresponding to the
quantum superposition principle. It can be given as follows.

For probabilities $p_1,p_2,p_3$,${\cal P}_1$, ${\cal P}_2$, ${\cal
P}_3$,$\Pi_1$, $\Pi_2$, and $\Pi_3$, the probabilities corresponding
to the density matrix $\rho_\psi$ are given by the expressions
\begin{eqnarray}
P_3 & = &({1}/{{\cal T}})\left\{\Pi_3p_3+(1-\Pi_3){\cal P}_3
  +2\sqrt{p_3{\cal P}_3}\left(\Pi_1-1/2\right)\right\},  \label{S20}\\
P_1-1/2& = &({1}/{{\cal T}})\Big\{\Pi_3(p_1-1/2)+({\cal P}_1-1/2)(1-\Pi_3)\nonumber\\
  &&+\left[(\Pi_1-1/2)(p_1-1/2)+(\Pi_2-1/2)(p_2-1/2)\right]\sqrt{{{\cal P}_3}/{p_3}}\nonumber\\
  &&+\left.\left[(\Pi_1-1/2)({\cal P}_1-1/2)-(\Pi_2-1/2)({\cal P}_2-1/2)\right]
    \sqrt{{p_3}/{{\cal P}_3}}\right\},\label{S21}\\
P_2-1/2 &= &({1}/{{\cal T}})\Big\{\left[(p_2-1/2)\Pi_3+({\cal P}_2-1/2)(1-\Pi_3)\right]\nonumber\\
  &&+\sqrt{{{\cal P}_3}/{p_3}}\left[(\Pi_1-1/2)(p_2-1/2)-(\Pi_2-1/2)(p_1-1/2)\right]\nonumber\\
  &&+\left.\sqrt{{p_3}/{{\cal P}_3}}\left[(\Pi_2-1/2)({\cal P}_1-1/2)
  +(\Pi_1-1/2)({\cal P}_2-1/2)\right]\right\},\label{S22}
\end{eqnarray}
where the parameter ${\cal T}$ reads
\begin{eqnarray}    
{\cal T} &=& 1+\frac{2}{\sqrt{p_3{\cal P}_3}}
  \left\{(\Pi_1-1/2)\left[(p_1-1/2)({\cal P}_1-1/2)
   +({\cal P}_2-1/2)(p_2-1/2)+p_3{\cal P}_3\right]\right.\nonumber\\
&&+\left.(\Pi_2-1/2)\left[(p_2-1/2)({\cal P}_1-1/2)-(p_1-1/2)
    ({\cal P}_2-1/2)\right]\right\},\label{S19}
\end{eqnarray}
and, in view of (\ref{eq.5a}), the probability $P_3$ can be
expressed through $P_2$ and $P_1$.

\section{Kinetic Equation for Qubits}   
The probabilities $p_1,$ $p_2$, and $p_3$ satisfy the evolution
equation, which follows from the von Neumann equation; in matrix
form, it reads
\begin{equation} \label{eq.A1}
\left(\begin{array}{cc}
  \dot p_3 & \dot p\\
  \dot p^\ast & -\dot p_3\end{array}\right)+
i\left[\left(\begin{array}{cc}
  z_1 & x-iy\\
  x+iy & z_2\end{array}\right)
  \!\!\!\!\begin{array}{cc}
  ~\\
  ,\end{array}\!\!\!\!\left(\begin{array}{cc}
  p_3 & p\\
  p^\ast & 1-p_3\end{array}\right)\right]=0,
\end{equation}
where $ p = p_1-(1/2)-i(p_2-1/2). $ This equation can be rewritten
as a linear equation for the complex vector $\vec p=(p_3,p,p^\ast)$;
we have
\begin{eqnarray}
&&i\frac{d p_3}{d t} = (x-i y)p^\ast-(x-i y)p,\quad
  i\frac{d p}{d t} = [z_1-z_2-2(x-i y)]p,\nonumber\\
&&-i\frac{d p^\ast}{d t} = [z_1-z_2-2(x+i y)]p^\ast.
\label{eq.A2}
\end{eqnarray}
Thus, the equation for the evolution of spin-1/2 system is mapped
onto the system of linear kinetic equations for the probabilities
identified with the system state. Parameters determining this
evolution are the random variables $x$, $y$, $z_1$, and $z_2$. The
solution to kinetic equation~(\ref{eq.A1}) or (\ref{eq.A2}) reads
\begin{equation} \label{eq.A3}
\left(\begin{array}{c}
  p_3(t)\\
  p(t)\\
  p^\ast(t)\\
  1-p_3(t)\end{array}\right)=u(t)\otimes u^\ast(t)
\left(\begin{array}{c}
  p_3(0)\\
  p(0)\\
  p^\ast(0)\\
  1-p_3(0)\end{array}\right),
\end{equation}
where the unitary 4$\times$4~matrix $u(t)\otimes u^\ast(t)$ is
expressed in terms of random variables using the unitary
2$\times$2~matrix $u(t)$ of the form
\begin{equation} \label{eq.A4}
u(t) = \exp\left[-i t\left(\begin{array}{cc}
  z_1 & x-i y \\
  x+i y & z_2\end{array}\right)\right].
\end{equation}
Thus, the unitary evolution of density matrix is expressed in terms
of the dependence of the probabilities on time corresponding to the
solution of kinetic equation~(\ref{eq.A2}).

\section{Quantum Suprematism Picture of Qubit States}   
In this section, we describe for qubit states the map of Bloch
sphere parameters $-1\leq a,b,c\leq 1$ onto the probabilities
$p_1,p_2$, and $p_3$ determining the density matrix $\rho$. The
Bloch sphere parameters are expressed in terms of the probabilities
as follows:
\begin{equation} \label{eq.X1}
a = 2p_1-1, \qquad b = 2p_2-1, \qquad c = 2p_2-1.
\end{equation}
They have the physical meaning of mean values of the spin projection
onto directions $x$, $y$, and $z$, i.e.,
\begin{equation} \label{eq.X2}
a = \frac{1}{2}\mbox{Tr}\rho\sigma_x, \qquad b=\frac{1}{2}\mbox{Tr}\rho\sigma_y,
  \qquad c = \frac{1}{2}\mbox{Tr}\rho\sigma_z,
\end{equation}

\begin{figure}
\vspace{-3mm}
\includegraphics[width=80mm]{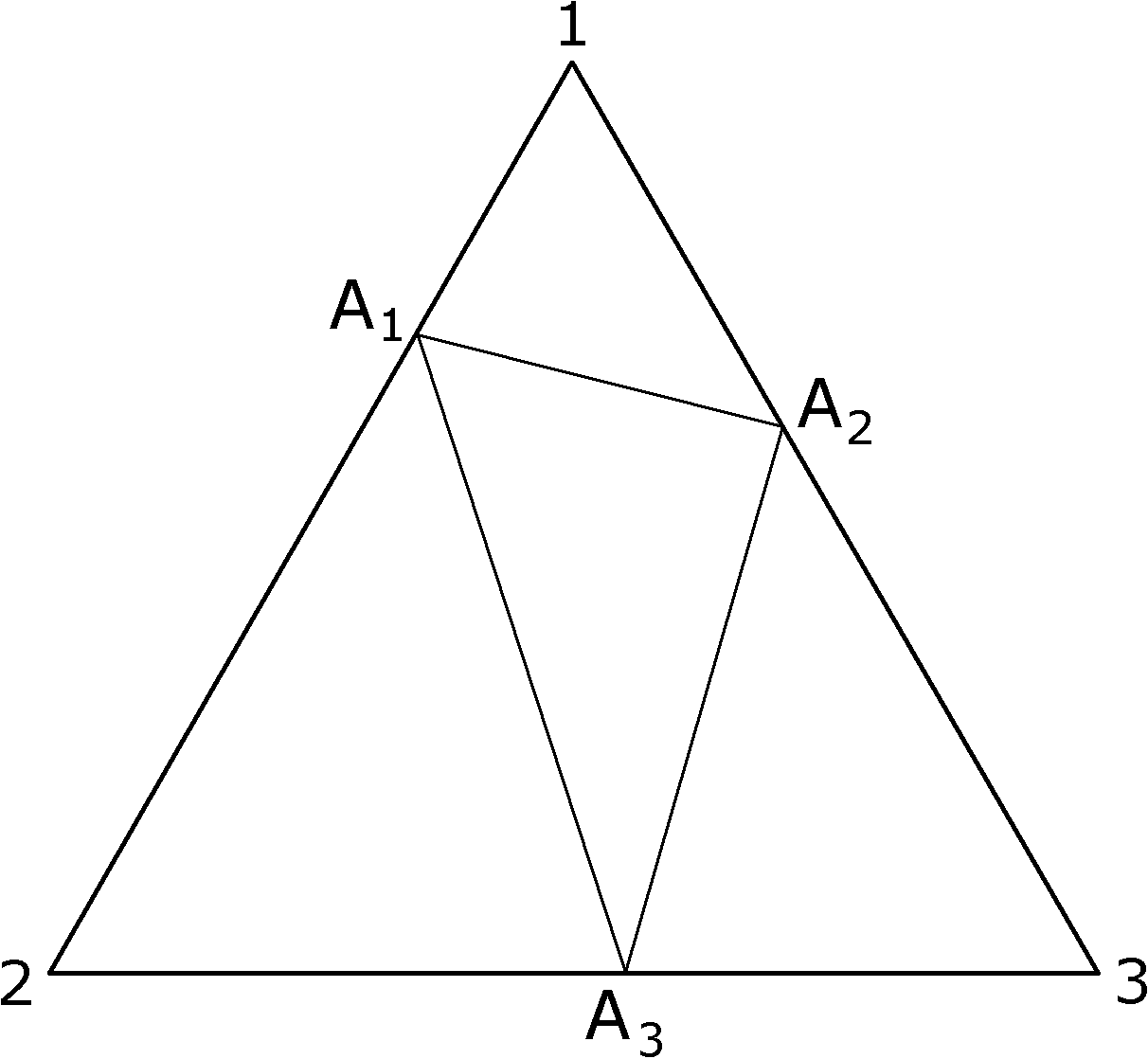}
\vspace{-3mm}
\caption{Triangle geometry of the qubit state.}
\end{figure}

\noindent  and satisfy the inequality
\begin{equation} \label{eq.X3}
a^2+b^2+c^2\leq 1.
\end{equation}
The points with coordinates $a,b$, and $c$ are situated in the Bloch
ball, and the points on the surface of Bloch sphere correspond to
the pure states of qubits. One can illustrate the qubit states using
triangle geometry and the triada of Malevich's squares. The triangle
$A_1A_2A_3$ inside the equilateral triangle with side equal to
$\sqrt2$ is shown in Fig.~1.

The vertices of the triangle are determined by the probabilities
$p_1$, $p_2$, and $p_3$~\cite{Chernega1,Chernega2}. The
correspondence is related to the bijective map of the Bloch
parameters and the probabilities~(\ref{eq.X1}). This map is
illustrated by three squares (black, red, and white) shown in Fig.~2
and constructed using the sides of triangle $A_1A_2A_3$. The sum of
the areas of the squares is expressed in terms of the probabilities
\begin{equation} \label{eq.X4}
S = 2[3(1-p_1-p_2-p_3)+2p_1^2+2p_2^2+2p_3^2+p_1p_2+p_2p_3+p_3p_1].
\end{equation}

\begin{figure}
\includegraphics[width=80mm]{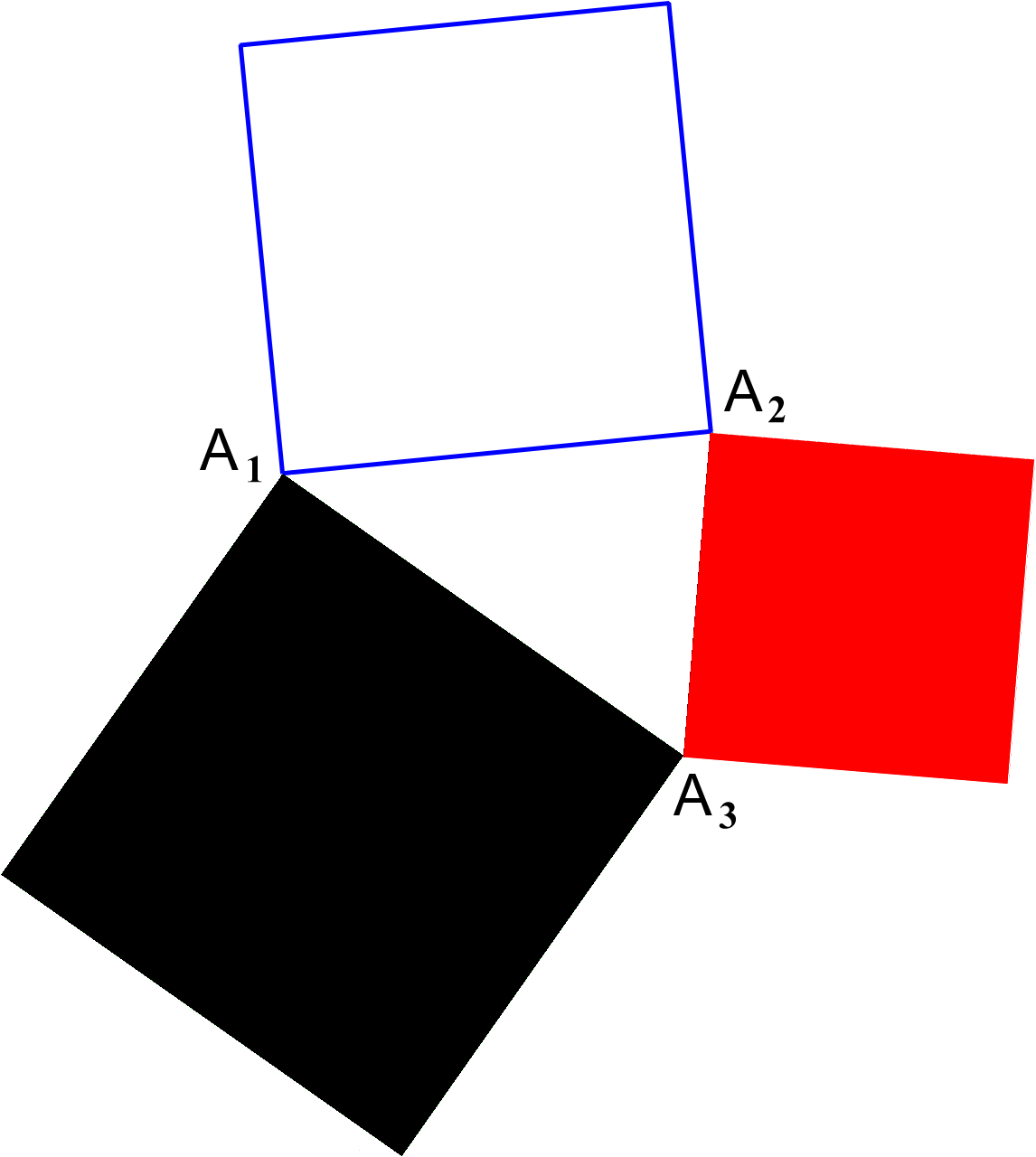}
\vspace{-3mm}
\caption{Triada of Malevich's squares.}
\end{figure}

\begin{figure}
\includegraphics[width=80mm]{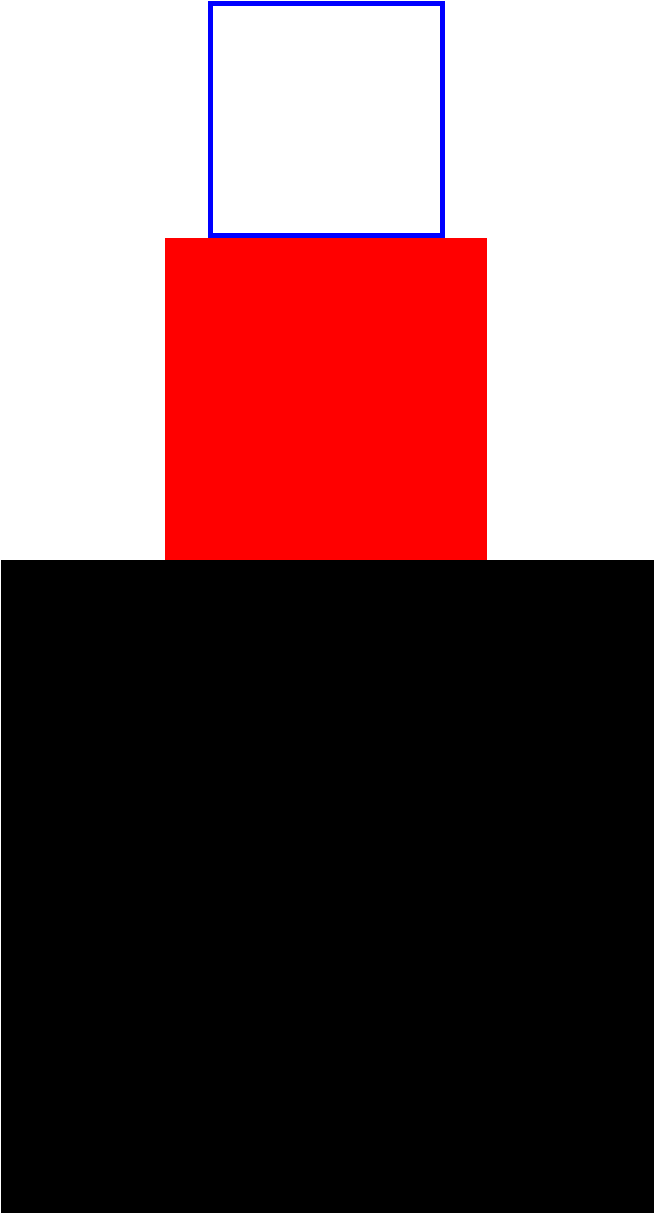}
\vspace{-3mm}
\caption{Three squares with different areas organized in
the form of Malevich's ``tower'' constructed of the three squares.}
\end{figure}

Thus, the point inside the Bloch ball is mapped onto the triada of
squares called Malevich's squares (see, Fig.2). The squares can be organized in
the form of Malevich's ``tower'' constructed of three squares; see
Fig.~3. The maximum value of area $S$ for classical coin statistics
is equal to $6$, which corresponds to the form of triangle
$A_1A_2A_3$ coinciding with the equilateral triangle with the side
$\sqrt 2$. The quantization condition provides the maximum value of
area $S=3$, which corresponds to the length of the side of the
equilateral triangle $A_1A_2A_3$ equal to one. The approach to
interpret the geometry of qubit states using the triangle and
Malevich's squares geometry was called the quantum suprematism
picture of qubit states in~\cite{Chernega1,Chernega2}. Thus, the
quantization procedure suggested to connect classical coin
statistics with quantum statistics of qubit states has a geometrical
interpretation in terms of the triangle geometry and the geometry of
the triada of Malevich's squares. The quantization provides the
prediction of quantum-to-classical ratio of areas $S_{\rm
classical}/S_{\rm quantum}=6/3=2$. It can be checked experimentally
measuring $p_1$, $p_2$, and $p_3$.

\section{Conclusions}       
To conclude, we point out our main results.

We presented a review of a new approach to the formulation of
quantum system states and observables, where the states are
described by probabilities. We show that the qubit state can be
given as a set of three probability distributions $(p_1,1-p_1)$,
$(p_2,1-p_2)$, and $(p_3,1-p_3)$, describing three classical coins.
We presented the superposition principle of the pure qubit states in
the form of the addition rule of these probability distributions.

The quantization procedure of classical coin tossing game expressed
as the condition for the coin probability distributions is
suggested. The quantum observables (Hermitian matrices) are
interpreted in terms of classical random variables. The addition
rule for the probabilities can be illustrated as a rule of the
combination of two triadas of Malevich's squares. The extension of
the obtained nonlinear addition rule for the probabilities
determining the superpositions of arbitrary qudit states can be
found.

One can illustrate the obtained bijective map of quantum states to
the classical coin probabilities using the statement of Albert
Einstein in the discussion with Niels Bohr on foundations of quantum
mechanics ``God does not play dice'' and replacing it with the
statement ``God plays coins.''

\end{document}